\documentclass[12pt,preprint]{aastex}

\slugcomment{To appear in the Astrophysical Journal}

\lefthead{Skinner et al.}
\righthead{LkCa 15}

\begin{document}

\title{XMM-Newton X-ray Observations of LkCa 15: A T Tauri Star With a Formative
       Planetary System}

\author{Stephen L. Skinner\footnote{CASA, Univ. of Colorado,
Boulder, CO, USA 80309-0389; stephen.skinner@colorado.edu} and
Manuel  G\"{u}del\footnote{Dept. of Astronomy, Univ. of Vienna, 
T\"{u}rkenschanzstr. 17,  A-1180 Vienna, Austria; manuel.guedel@univie.ac.at}}

%
\newcommand{\ltsimeq}{\raisebox{-0.6ex}{$\,\stackrel{\raisebox{-.2ex}%
{$\textstyle<$}}{\sim}\,$}}
%
\newcommand{\gtsimeq}{\raisebox{-0.6ex}{$\,\stackrel{\raisebox{-.2ex}%
{$\textstyle>$}}{\sim}\,$}}

\begin{abstract}
High-resolution ground-based images of the T Tauri star LkCa 15 have
revealed multiple companions that are thought to comprise a formative
planetary system. The candidate protoplanets orbit at distances   $\sim$15 - 20 AU 
within the dust-depleted inner region of the circumstellar disk. Because 
of its young age ($\sim$1 - 4 Myr), LkCa 15 provides a benchmark system 
for testing planet-formation models. We  detected LkCa 15 as 
a bright X-ray source in a short 10 ks {\em Chandra} observation in  
2009. We report here new results obtained from a deeper 37 ks  {\em XMM-Newton} 
observation  in  2014. The new data provide better sampling in the time domain 
and improved sensitivity at low energies below 1 keV. 
Spectral fits with thermal emission models require at least two temperature
components at kT$_{cool}$ $\approx$ 0.4 keV and kT$_{hot}$ $\approx$ 2.2 keV.
The value of kT$_{hot}$ is about a factor of two less than inferred from
{\em Chandra}, suggesting that the hot-component temperature 
is variable. The best-fit absorption column density is in good agreement with 
that expected from optical extinction estimates A$_{\rm V}$ $\approx$ 1.3 - 1.7 mag.
The intrinsic X-ray luminosity is L$_{x}$(0.2 - 10 keV) = 3 $\times$ 10$^{30}$ ergs s$^{-1}$.
Estimates of the X-ray heating rate of the inner disk and protoplanets are sensitive to
the assumed disk gas surface density for which recent  ALMA observations give estimates 
$\Sigma_{0,gas}$ $\sim$ 10$^{2}$ g cm$^{-2}$ at 1 AU from the star. At such 
densities, X-ray heating is confined mainly to the upper disk layers  
and X-ray penetration through the disk midplane to the 
protoplanets at $r$ $\approx$ 15 - 20 AU is negligible.
\end{abstract}


\keywords{stars: individual (LkCa 15; NSVS 6777197) --- accretion, accretion disks --- 
stars: pre-main sequence --- X-rays: stars}



\section{Introduction}

Although several thousand  exoplanets have now been 
discovered\footnote{For a current catalog  of  exoplanets, see the
{\em NASA Exoplanet Archive} database at
http://http://exoplanetarchive.ipac.caltech.edu.},  examples of
exoplanets orbiting young pre-main sequence (PMS) stars are rare.
The identification of planet-hosting young stars of ages a few Myr  
and study of their circumstellar disks provide valuable insight into 
the circumstellar environment in which planets (and planetary systems) 
form and how the planets affect disk properties such as gas and dust
distribution. As such, observational studies of disks and exoplanets 
around PMS host stars provide crucial constraints on 
planet-formation models.

Perhaps the most striking example to emerge so far of a PMS star hosting
a protoplanet, and possibly even a protoplanetary system, is the accreting 
classical T Tauri star (cTTS) LkCa 15 in the Taurus star-forming region (Table 1). 
In a remarkable discovery, Kraus \& Ireland (2012, hereafter KI12) reported 
the direct detection using infrared masked aperture interferometry of a suspected 
protoplanet at a projected separation of 71.9 $\pm$ 1.6 mas from LkCa 15.
The protoplanet is located in the dust-depleted inner region of the LkCa 15
disk at a deprojected orbital radius of $\sim$16 - 20 AU.
Further monitoring in the near-IR has detected possible orbital motion of the 
protoplanet (Ireland \& Kraus 2014).

Extensive and ongoing studies of the disk surrounding the host star LkCa 15
at infrared, (sub)millimeter, and radio wavelengths show that it is
severely depleted of dust inside a radius of $\sim$45 - 50 AU 
(Andrews et al 2011a,b - hereafter A11a,b; Isella et al. 2012, 2014; 
Espaillat et al. 2008, 2010; Thalmann et al. 2014, 2016). 
Although the inner region is dust-depleted,
there is still gas present as revealed by $^{12}$CO and $^{13}$CO 
observations (Pi\'{e}tu et al. 2007; van der Marel et al. 2015). 
Also, Isella et al. (2014) detected
a compact 7 mm continuum source with the Very Large Array (VLA) at the 
position of the central star. They conclude that the 7 mm emission is
not consistent with a stellar photospheric origin but could be due
to either millimeter size grains near the star or ionized gas in
the vicinity of the star. Plausible explanations for the dust-clearing in
the inner disk are variable dust grain sizes and opacity (Isella et al. 2012)
or dynamical clearing by one or more orbiting objects (A11a,b), with 
the latter explanation currently favored.

The possibility that more than one protoplanet might be present has
received some recent support from  new high-contrast near-IR imaging
and adaptive optics H$\alpha$ imaging  of LkCa 15 obtained by 
Sallum et al. (2015). They report the detection of three 
distinct objects with best-fit orbital semi-major axes in the range
14.7 - 18.6 AU. One of these objects is the suspected protoplanet 
that was designated as LkCa 15b by KI12 but is referred to as LkCa 15c
by Sallum et al. (2015). The latter authors obtain a semi-major axis 
$a$ = 18.6 ($+$2.5, $-$2.7) AU for this object. They also report the 
direct near-IR and H$\alpha$ detection of an object orbiting closer to the 
star at $a$ = 14.7 ($\pm$2.1) AU which they characterize  as an 
accreting protoplanet. A third fainter object at $a$ $\sim$ 18 AU
was detected at L$'$-band only and its properties are not yet well-constrained.
Studies of the inner disk and candidate protoplanets are ongoing.
High spatial resolution scattered light images of the inner disk of
LkCa 15 based on J-band imaging polarimetry have now been obtained by
Thalmann et al. (2016). Their images show structure from scattering
material at the positions of the candidate protoplanets which they
argue could be  responsible for {\em some} of the signal reported
in previous aperture-masking observations.

The likely presence of a formative planetary system around LkCa 15
provides an unprecedented opportunity to study the early stages of
planet formation in detail  at high effective spatial resolution
given the modest distance of $\sim$140 pc to the Taurus star-forming 
region. Our study presented  here focuses on the effects of 
X-ray irradiation by the central star on the gas-dominated inner disk.
We detected LkCa 15 as a bright X-ray source in a previous {\em Chandra} 
observation and provided initial estimates of the X-ray ionization and heating
rates in the inner disk (Skinner \& G\"{u}del 2013, hereafter SG13).

We report here the results of a more recent X-ray observation of
LkCa 15 obtained with {\em XMM-Newton}. These observations provide
better sensitivity at low energies below 1 keV where X-ray absorption
due to intervening gas becomes important. The improved low-energy
sensitivity provides tighter constraints on the absorption toward
the star as measured by the equivalent neutral hydrogen column 
density N$_{\rm H}$, and the intrinsic (unabsorbed) X-ray
luminosity (L$_{x}$). Fits of the {\em XMM-Newton} spectra 
confirm the earlier {\em Chandra} result that cool and hot plasma 
components are present but the temperature of  the 
hot component is lower than was measured with  {\em Chandra}
and variability of the hotter plasma seems likely. 
X-ray ionization and heating rates of the inner disk are 
recomputed based on the new {\em XMM-Newton} results
and revised inner disk model parameters from recent
ALMA observations.

The importance of the effects of  X-ray and extreme-ultraviolet (EUV) radiation 
on disks and protoplanetary systems around young stars has been noted in several
previous studies of which  two examples are Igea \& Glassgold (1999; hereafter IG99) 
and Cecchi-Pestellini, Ciaravella, \& Micela (2006). X-ray and EUV
emission ionizes and heats disk gas (especially in the  outer surface layers),
affects disk chemistry, accretion, and mass-loss (via photoevaporation), and 
strengthens the coupling between the accretion disk and the stellar magnetic field.
Since X-rays influence mass-loss and disk dissipation they are one of the 
factors that constrain the timescale for planet formation.

\clearpage                                                                                                                                              
\begin{deluxetable}{lll}
\tabletypesize{\footnotesize}
\tablewidth{0pt}
\tablecaption{Properties of LkCa 15}
\tablehead{
           \colhead{Property}               &
           \colhead{Value}                  &
           \colhead{Refs.}                  
}
\startdata
Sp. type                    & K5$\pm$2        &  1,2  \nl
Age (Myr)                   & 2 [1 - 4]         &  3    \nl
M$_{*}$ (M$_{\odot}$)       & 1.0             &  4  \nl
R$_{*}$ (R$_{\odot}$)       & 1.6             &  2  \nl
T$_{eff}$ (K)               & 4730            &  1  \nl
A$_{\rm V}$ (mag)           & 0.62; 1.3 - 1.7 &  1,2  \nl
$i_{disk}$ (deg.)           & 50.5 - 52       &  5 \nl
M$_{disk}$ (M$_{\odot}$)    & $\approx$0.1    &  5,6  \nl
d (pc)                      & 140             &  7  \nl
L$_{*}$ (L$_{\odot}$)       & 0.74 - 1.2      &  4,8   \nl
log L$_{x}$ (ergs s$^{-1}$) & 30.47 [30.40-30.52]   &  9,10    \nl
\enddata
\tablecomments{Refs.~
(1) Kenyon \& Hartmann 1995~
(2) Espaillat et al. 2010~
(3) Kraus \& Hillenbrand 2009~
(4) Simon et al. 2000~
(5) Isella et al. 2012~
(6) van der Marel et al. 2015~
(7) Torres et al. 2009~
(8) Andrews et al. 2011b~
(9) Skinner \& G\"{u}del 2013~
(10) this work 
}
\end{deluxetable}
\clearpage


\section{Previous Chandra Observation}

We obtained a  9.8 ks {\em Chandra}  observation (ObsId 10999) of
LkCa 15  on 27 December 2009 using the ACIS-S (Advanced CCD
Imaging Spectrometer) array. Results were presented by 
SG13 and are briefly
summarized here. LkCa 15 was detected with 590 net counts (0.2 - 8 keV).
No statistically significant variability was present in the X-ray
light curve. Acceptable spectral fits were obtained using a two-temperature (2T)
$apec$ thermal plasma model and abundances typical of TTS in Taurus.
The best-fit model gave an absorption column density 
N$_{\rm H}$ = 3.7 [2.4-5.1] $\times$ 10$^{21}$ cm$^{-2}$ and plasma
temperatures kT$_{cool}$ = 0.30 [0.25 - 0.37 keV], 
kT$_{hot}$ = 5.1 [3.0 - 13.6] keV, where brackets enclose
90\% confidence intervals. The intrinsic (unabsorbed) luminosity was
log L$_{x}$(0.3 - 10 keV) = 30.4 ergs s$^{-1}$ at an assumed distance
of 140 pc.

\section{XMM-Newton  Observation}

The {\em XMM-Newton} observation (ObsId 0722340101) began on  
20 February 2014 at 19:12 UTC and ended on 21 February at
06:02 UTC.  Data were acquired with the European Photon
Imaging Camera (EPIC) in Full-Frame mode using the 
Medium optical blocking filter.  EPIC provides
charge-coupled device (CCD) imaging spectroscopy from the
pn camera (Str\"{u}der et al. 2001) and two nearly
identical MOS cameras (MOS1 and MOS2; Turner et al. 2001).
The EPIC cameras have  energy coverage in the range
E $\approx$ 0.2 - 15 keV with energy
resolution E/$\Delta$E $\approx$ 20 - 50.
The MOS cameras provide the best on-axis angular
resolution with FWHM $\approx$ 4.3$''$ at 1.5 keV.

Data were reduced with the {\em XMM-Newton}
Science Analysis System (SAS vers. 15.0) using
standard procedures including the filtering of
raw event data to select good event patterns
and removal of data within time intervals of
high background radiation. The usable exposures
obtained after removing high background intervals
and total exposures (in parentheses) were
27.8 (36.04) ks for pn, 35.8 (37.70) ks for MOS1,
and 36.6 (37.67) ks for MOS2. Thus, about 22\% of
the pn exposure was adversely affected by high
background but the MOS exposures were not 
severely affected.

A circular region centered on LkCa 15 with a 
radius r = 20$''$ ($\approx$80\% encircled energy at 1.5 keV)
was used to extract X-ray light curves and spectra.
Background analysis was conducted on circular source-free regions near
the source. The SAS tasks {\em rmfgen} and {\em arfgen} were
used to generate source-specific RMFs and ARFs for spectral
analysis. The data were analyzed using the HEASOFT {\em Xanadu}
software package.

\section{Results}

Figure 1 shows the EPIC pn image of LkCa 15 and its surroundings.
LkCa 15 is prominently detected at pn centroid position 
R.A. = 04$^{h}$ 39$^{m}$ 17$^{s}$.79, decl. = $+$22$^{\circ}$ 21$'$ 03$''$.08 (J2000).
The previous {\em Chandra} observation had somewhat better spatial resolution 
and gave an X-ray position R.A. = 04$^{h}$ 39$^{m}$ 17$^{s}$.787, 
decl. = $+$22$^{\circ}$ 21$'$ 03$''$.28 (SG13). These X-ray positions are
in good agreement with the {\em HST} GSC v2.3 position
R.A. = 04$^{h}$ 39$^{m}$ 17$^{s}$.787, decl. = $+$22$^{\circ}$ 21$'$ 03$''$.26.
Evidence for binarity in LkCa 15 has so far not been found (Nguyen et al. 2012).

There are no other X-ray sources in the immediate vicinity of LkCa 15.
The nearest EPIC source detected by the pipeline processing
software lies more than 1$'$ from LkCa 15.
The K5 star HD 284589 located 27.$''$6 north of LkCa 15 is not detected,
nor was it detected by {\em Chandra}. The only other  bright stellar X-ray source 
in the EPIC field-of-view is the eclipsing Algol-type binary system
NSVS 6777197 (2MASS J04394628$+$2211503) located $\approx$11$'$ southeast of
LkCa 15 (Fig. 1). Its X-ray properties are summarized further below.

The EPIC light curves of LkCa 15 are shown in Figure 2.  No large-amplitude fluctuations
or flares are present but there is  a slow falloff in the count
rate during the observation. Checks for variability  on binned background-subtracted
broad-band (0.2 - 8 keV) light curves using the $\chi^2$ test give a high variability probability.
Using 1000 s bins, the probability of variability is
P$_{var}$(0.2 - 8 keV) = 0.96 (pn), 0.99 (MOS1), and 0.98 (MOS2). 

Figure 3 compares the EPIC pn spectrum with the previous {\em Chandra} ACIS-S 
spectrum and shows an overlay of the two EPIC MOS spectra.
A notable difference is that there is no significant emission below
$\approx$0.5 keV in the ACIS-S spectrum but the higher  EPIC pn
effective area at low energies reveals detectable emission down to 
$\approx$0.2 keV. This leads to a  more reliable determination
of N$_{\rm H}$ from the EPIC pn spectrum since lower-energy photons are more  
susceptible to absorption. The only line feature clearly visible is the 
Ne X Ly$\alpha$ line at 1.02 keV which is seen in both MOS spectra.
There is also a weak feature at 1.86 keV in the pn spectrum that may
be Si XIII.

Spectra were fitted using the Astrophysical Plasma Emission Code
variable-abundance {\em vapec} model in XSPEC version 12.8.2 (Smith et al. 2001).
Photoelectric absorption was applied using the XSPEC {\em wabs} model to
determine the equivalent hydrogen column density (N$_{\rm H}$).
A two-temperature (2T) {\em vapec} model was required  to obtain
acceptable fits. A comparison of fit results for two {\em vapec} models
using different abundances is given in Table 2.
Significant improvement relative to the solar-abundance
fit is obtained by allowing the abundances of Ne and Fe to vary
(model A in Table 2). Very little further improvement in the
fit is obtained by letting the abundances of other metals to 
deviate from solar values.  Model B uses typical TTS abundances
for the Taurus Molecular Cloud (G\"{u}del et al. 2007; Scelsi et al. 2007), 
as were adopted in the {\em Chandra} analysis (Model C of SG13).
The Ne abundance inferred from Model A is Ne = 2.4 [2.07 - 2.71]
times solar, but the generic Taurus abundances (Model B) keep
the value fixed at Ne = 0.83 times solar.

The EPIC fit results for Models A and B are overall quite similar.
Cool plasma at kT$_{1}$ $\approx$ 0.4 keV is required by both models,
consistent with the value obtained from previous {\em Chandra} ACIS-S
data (SG13). A hotter component at  kT$_{2}$ $\approx$ 2.1 - 2.4  keV
is also required. This temperature is about a factor of two lower than was 
inferred from {\em Chandra}, which gave kT$_{2}$ $\approx$ 4 - 5 keV.
The 90\% confidence levels  from EPIC are kT$_{2}$ = [1.93 - 2.40] keV
(model B), which does not overlap the corresponding ACIS-S
range kT$_{2}$ = [3.0 - 13.6] keV. This suggests that the temperature of
the hot component is variable. 
Model B associates a larger percentage of the volume emission
measure with the cool plasma component, as determined by the 
XSPEC normalization parameter ($norm$).

The best-fit absorption column density is consistent with 
that expected from A$_{\rm V}$ = 1.3 - 1.7 (Table 1; Espaillat et al. 2010).
Using the N$_{\rm H}$ to A$_{\rm V}$ conversion of
Vuong et al. (2003), model A gives
A$_{\rm V}$ = 1.4 [1.3 - 1.6]  and model B yields
A$_{\rm V}$ = 1.8 [1.7 - 1.9]. The Gorenstein (1975)
coversion gives A$_{\rm V}$ values that are about
25\% lower than above. Thus, there is no convincing
evidence for excess X-ray absorption above that
expected from A$_{\rm V}$. 
The ratio of X-ray to stellar luminosity from the EPIC fits is
log L$_{x}$(0.2 - 10 keV)/L$_{*}$ = $-$3.16 to $-$3.06. This ratio
is in good agreement with other TTS in Taurus as determined by Telleschi et al. (2007).

{\em NSVS 6777197}:~The star NSVS 6777197  was serendipitously detected as a
bright X-ray source by EPIC pn (Fig. 1) and MOS2, but was outside 
the MOS1 field-of-view.
It was classified as a detached Algol binary with a 3.928 d period
by Drake et al. (2014). Algol-type binaries usually have one
component of late spectral type and are often detected as
bright coronal X-ray sources (Singh, Drake, \& White 1996). 
Large X-ray flares can occur as seen in the prototype
Algol (Schmitt \& Favata 1999).
But the EPIC X-ray light curves of NSVS 6777197 show no
significant variability.  Simultaneous spectral fits of the 
pn and MOS2 spectra with a 2T
$vapec$ thermal plasma model give an absorption
column density N$_{\rm H}$ = 1.78 [1.57 - 2.04] $\times$ 10$^{21}$ cm$^{-2}$,
kT$_{cool}$ = 0.92 [0.79 - 1.04] keV, kT$_{hot}$ = 2.60 [2.23 - 3.26] keV,
and iron abundance Fe = 0.22 [0.10 - 0.36] solar,
where brackets enclose 90\% confidence intervals.
The absorbed (and unabsorbed) fluxes are F$_{x}$(0.2-10 keV) =
7.2 $\times$ 10$^{-13}$ (1.10 $\times$ 10$^{-12}$) ergs cm$^{-2}$ s$^{-1}$. 

\clearpage                                                                                                                                     
\begin{deluxetable}{lll}
\tabletypesize{\scriptsize}
                                                                                                                                     
\tablewidth{0pc}
\tablecaption{{\em XMM-Newton} Spectral Fits for LkCa 15
   \label{tbl-1}}
\tablehead{
\colhead{Parameter}      &
\colhead{ }              &
\colhead{  }
}
\startdata
Model                                   &    A                        & B               \nl
Emission\tablenotemark{a}               &  Thermal (2T)               & Thermal (2T)    \nl
Abundances                              &  non-solar\tablenotemark{c} & non-solar\tablenotemark{d}    \nl
N$_{\rm H}$ (10$^{22}$ cm$^{-2}$)       & 0.23 [0.21 - 0.25]          & 0.29 [0.27 - 0.31] \nl
kT$_{1}$ (keV)                          & 0.41 [0.39 - 0.44]          & 0.40 [0.39 - 0.42] \nl
kT$_{2}$ (keV)                          & 2.39 [2.20 - 2.59]          & 2.08 [1.93 - 2.40] \nl
norm$_{1}$ (10$^{-4}$)\tablenotemark{b} & 2.75 [2.31 - 3.29]          & 8.26 [7.33 - 9.28] \nl
norm$_{2}$  (10$^{-4}$)\tablenotemark{b}& 3.06 [2.91 - 3.21]          & 4.22 [3.84 - 4.50] \nl
$\chi^2$/dof                            & 862.0/634                   & 850.4/636             \nl
$\chi^2_{red}$                          & 1.36                        & 1.34                \nl
F$_{\rm X}$ (10$^{-12}$ ergs cm$^{-2}$ s$^{-1}$)   & 0.55 (1.13)                 & 0.54 (1.43)         \nl
F$_{\rm X,1}$ (10$^{-12}$ ergs cm$^{-2}$ s$^{-1}$) & 0.23 (0.64)                 & 0.24 (0.92)         \nl
F$_{\rm X,2}$ (10$^{-12}$ ergs cm$^{-2}$ s$^{-1}$) & 0.32 (0.49)                 & 0.29 (0.51)  \nl
log L$_{\rm X}$ (ergs s$^{-1}$)                    & 30.42                       & 30.52               \nl
log L$_{\rm X,1}$ (ergs s$^{-1}$)                  & 30.18                       & 30.33             \nl
log L$_{\rm X,2}$ (ergs s$^{-1}$)                  & 30.06                       & 30.08               \nl
log [L$_{\rm X}$/L$_{*}$]                        & $-$3.16                     & $-$3.06             \nl
\enddata
\tablecomments{
Based on  XSPEC (version 12.8.2) simultaneous fits of the background-subtracted EPIC spectra
(pn, MOS1, MOS2) binned to a minimum of 10 counts per bin with high background time intervals removed.
The fits were obtained with 2T $vapec$ optically plasma models. The tabulated parameters
are absorption column density (N$_{\rm H}$), plasma energy (kT),
and XSPEC component normalization (norm).
Abundances are referenced to the solar values of  Anders \& Grevesse (1989).
Square brackets enclose 90\% confidence intervals.
The total X-ray flux (F$_{\rm X}$) and fluxes associated with each model component
(F$_{\rm X,i}$)  are the absorbed values in the 0.2 - 10 keV range, followed in
parentheses by  unabsorbed values.
The total X-ray luminosity L$_{\rm X}$  and luminosities of each component
L$_{\rm X,i}$ are unabsorbed values in the 0.2 - 10 keV range and  assume a
distance of 140 pc. A value L$_{*}$ = 1.0 L$_{\odot}$ is adopted
based on an average of values given in the literature.}
\tablenotetext{a}{Models A and B are of form:~N$_{\rm H}$$\cdot$(kT$_{1}$ $+$ kT$_{2}$)} \\
\tablenotetext{b}{For thermal $apec$ models, the norm is related to the volume emission measure
                  (EM = n$_{e}^{2}$V)  by
                  EM = 4$\pi$10$^{14}$d$_{cm}^2$$\times$norm, where d$_{cm}$ is the stellar
                  distance in cm. At d = 140 pc this becomes
                  EM = 2.34$\times$10$^{56}$ $\times$ norm (cm$^{-3}$). }
\tablenotetext{c}{All abundances were held fixed at their solar values except for Ne and Fe which were allowed to
                  vary and converged to Ne = 2.37 [2.07 - 2.71] and Fe = 0.45 [0.39 - 0.53] relative to their solar values.}
\tablenotetext{d}{Abundances were held fixed at typical values for TTS in Taurus (G\"{u}del et al. 2007; Scelsi et al. 2007).
                  These are (relative to solar): H = 1.0, He = 1.0, C = 0.45, N = 0.79, O = 0.43,
                   Ne = 0.83, Mg = 0.26, Al = 0.50, Si = 0.31, S = 0.42, Ar = 0.55, Ca = 0.195,
                   Fe = 0.195, Ni = 0.195. }
\end{deluxetable}
\clearpage

\section{Discussion}

\subsection{X-ray Heating and Ionization}
 
Disk X-ray heating is due to fast electrons ejected by
atoms during X-ray ionization.
A complete discussion of the methodology for computing the 
X-ray ionization rates ($\zeta$) and heating rates ($\Gamma_{x}$)
in the disk can be found in the {\em Chandra} study of 
LkCa 15 (SG13) and  the references  cited below. We have recomputed
the disk ionization and heating rates for LkCa 15 based on 
plasma temperature (kT$_{x}$) and X-ray luminosity (L$_{x}$)
values determined from the {\em XMM-Newton} spectral fits (Table 2).
The revised rates for the cool and hot plasma components
are summarized in Table 3.

\subsubsection{Disk Model}

A disk model must be adopted in order to
compute X-ray ionization and heating rates. A key disk 
parameter is the density of H-nuclei (n$_{\rm H}$),
which is required to compute the X-ray absorption and 
heating rate per unit volume at a  given position in the disk.
The value of n$_{\rm H}$ is determined by the disk gas
surface density $\Sigma$ which is not yet well-constrained 
observationally in the inner disk because of limitations on 
telescope angular resolution, especially at (sub)-mm wavelengths.

A cylindrical coordinate system is used to specify 
positions ($r$,$z$) in the disk where the radial coordinate
$r$ is the distance in the midplane from the center of the 
star to the specified point and $z$ is the height above
the disk midplane. Azimuthal symmetry is assumed.
We adopt a radial disk temperature profile of the form
T($r$) = 400($r$/1 AU)$^{-0.5}$ K based on a stellar effective
temperature T$_{eff}$ = 4730 K. Vertical temperature gradients are 
ignored (IG99). For the above temperature
relation and adopted stellar parameters (Table 1) the disk scale
height at $r$ = 1 AU is 
H$_{0}$ $\equiv$ H($r$ = 1 AU) = 7.1 $\times$ 10$^{11}$ cm 
and scales as H($r$) $\propto$ $r^{+1.25}$. 

We adopt a simple power-law for the gas surface density 
$\Sigma(r)$ = $\Sigma_{0}$($r$/1 AU)$^q$, normalized to 
$\Sigma_{0}$ = 10$^{2}$  g cm$^{-2}$ and a power-law exponent $q$ = $-$1. 
This simple power-law  ignores the additional exponential decay 
term included in the surface density profiles of A11b and van der Marel 
et al. (2015) since its value is near unity for the inner disk radii 
$r$ $\ltsimeq$ 15 AU of interest here.
The above value of $\Sigma_{0}$ is comparable to that determined from
the recent studies of Manara et al. (2014) and van der Marel et al. (2015)
but must be interpreted as  an order-of-magnitude estimate due to
several factors that limit our knowledge of the gas distribution 
in the spatially-unresolved inner disk (Sec. 5.2). We also note that the value 
of $\Sigma_{0}$ adopted here is much higher than the value  
$\Sigma_{0}$ = 10$^{-3}$ g cm$^{-2}$ used in SG13 that was based on the 
earlier LkCa 15 disk study of A11b (Sec. 5.2).

We adopt the abundance ratio  He/H = 0.1 by number and assume
that hydrogen in the disk is predominantly molecular. The mass
density at the midplane is $\rho$($r$,$z$=0) = 0.4$\Sigma(r)$/H$(r)$
which at $r$ = 1 AU gives $\rho_{0}$ $\equiv$  
$\rho$($r$=1 AU,$z$=0) = 5.63 $\times$ 10$^{-16}$ g cm$^{-3}$.
The number density of H-nuclei is given by n$_{\rm H}$ = $\rho$/($\mu$m$_{p}$) 
where m$_{p}$ is the proton mass and $\mu$ = 1.42 for H-nuclei
(Glassgold et al. 2004).

\subsubsection{Disk Ionization and Heating}

Calculation of the X-ray ionization and heating rates in the disk 
is based on the analytic results of 
Glassgold et al. (1997a, hereafter G97a;  Glassgold et al. 1997b), 
IG99, Shang et al. (2002; hereafter S02), and Glassgold et al. (2004).  
The X-ray emission is modeled as a thermal plasma with a 
characteristic temperature T$_{\rm x}$. The ionization 
rate scales as the inverse-square of the distance from the
source according to  (eq. [3.9] of S02):

\begin{equation}
\zeta \approx \zeta_{\rm x} \left[{ \frac{r}{R_{\rm x}}} \right]^{-2} \left[{ \frac{kT_{x}}{\epsilon_{ion}}} \right] I_{p}(\tau_{\rm x}, \xi_{0})~~~{\rm s}^{-1}~~{\rm (per~ H~nucleus)}~.
\end{equation}

\noindent In the above,  $R_{\rm x}$ fixes the height of the X-ray source above (or below)
the disk center to mimic X-ray production in coronal loops. We use
R$_{\rm x}$ = 4R$_{*}$ = 6.4 R$_{\odot}$ = 4.45 $\times$  10$^{11}$  cm
as in previous studies (G97a, S02, SG13) but 
the results are not very sensitive to the adopted value of R$_{\rm x}$ (IG99). 
The energy required to produce  an ion pair is $\epsilon_{ion}$ $\approx$ 37 eV.
The term $I_{p}(\tau_{\rm x}, \xi_{0})$ accounts for X-ray attenuation
as a function of optical depth $\tau_{\rm x}$, where the latter depends
on the photon energy $E$ and position in the disk.
The term  $\xi_{0}$ = E$_{0}$/kT$_{\rm x}$ 
applies a low-energy cutoff E$_{0}$ to the X-ray spectrum to account for wind 
absorption (IG99, S02). We use E$_{0}$ = 0.1 keV here, as in previous
work (S02, SG13). Figure 3 of SG13 shows the  effect of decreasing the cutoff 
energy to  E$_{0}$ = 0.01 keV.
 
The primary ionization rate $\zeta_{\rm x}$  is 
given by (S02):

\begin{equation}
\zeta_{\rm x} =  \frac{L_{x}\sigma(kT_{x})}{4 \pi R_{x}^2 kT_{x}}  =  1.13 \times 10^{-8}  \left[{ \frac{L_{x}}{10^{30}~ {\rm erg~ s}^{-1}}} \right] \left[{ \frac{kT_{\rm x}}{{\rm keV}}} \right]^{-(p+1)} \left[{ \frac{R_{\rm x}}{10^{12}~{\rm cm}}} \right]^{-2}~~~{\rm s}^{-1}
\end{equation}
where $\sigma(kT_{x})$ =  $\sigma(E)$  is the energy-dependent photoelectric X-ray 
absorption cross-section  per H nucleus. It is evaluated using the expression
$\sigma(E)$ = $\sigma_{0}$(E/1 keV)$^{-p}$ cm$^{-2}$ where 
$\sigma_{0}$ = 2.27 $\times$ 10$^{-22}$ cm$^{2}$ and
$p$ = 2.485 for solar-abundance disk plasma (G97a).
Numerical values of $\zeta_{\rm x}$ for the cool and  hot plasma components are given in 
Table 3 notes.

The X-ray optical depth used to compute the attenuation factor $I_{p}(\tau_{\rm x}, \xi_{0})$  is

\begin{equation}
\tau_{x}(r,z,E) =  \left[{ \frac{r}{R_{\rm x}}}  \right] \sigma({\rm E}) {\rm N}_{\perp,{\rm disk}}(r,z)
\end{equation}
 
where  the vertically-integrated column density from infinity
down to the  height z  above the disk midplane is 

\begin{equation}
{\rm N}_{\perp,{\rm disk}}(r,z) = \int^{\infty}_{z} {\rm n_{H}(r,\overline{z}) d\overline{z}}~~~{\rm cm^{-2}}~~.
\end{equation}

 In general, $\tau_{x}$ will be smaller for the 
hot component  at a given point in the disk, allowing the
harder emission from the hot component to penetrate deeper into the disk.
Note that for a given energy $E$ and height $z$, the optical depth $\tau_{x}$ is independent
of $r$ as the  result of a cancellation that occurs for the adopted
surface density profile $\Sigma(r)$ $\propto$ $r^{-1}$.
Also, it is worth emphasizing that the height above the midplane corresponding
to $\tau_{x}$ = 1 depends sensitively on the adopted value of the
gas surface density $\Sigma_{0}$ as shown in Figure 4.
At a given distance from the star, smaller adopted values of $\Sigma_{0}$ correspond
to lower scale heights and deeper X-ray penetration into the disk.
The ionization rate for the cool and hot components as a function of
$\tau_{x}$ is plotted in Figure 5.

The  X-ray heating rate per unit volume is (G12)

\begin{equation}
\Gamma_{\rm x} = \zeta {\rm n_{H} Q}
\end{equation}
where $Q$ is the heating rate per ionization. Several 
processes can affect the heating rate as discussed by G12.
For predominantly molecular disk gas at $r$ $\geq$ 1 AU
and the range of densities for the LkCa 15 inner disk
(Table 3) we adopt $Q$ = 17 eV = 2.72 $\times$ 10$^{-11}$ ergs,
as in previous studies (SG13, G12).
The X-ray heating rate then becomes

\begin{equation}
\Gamma_{\rm x} = 2.72 \times 10^{-11} \zeta  n_{\rm{H}}~~~~{\rm ergs~ s^{-1} cm^{-3}} 
\end{equation}

The heating rates at $r$ = 1 AU  given in Table 3
can be scaled to other radii using the scaling relations
for $\zeta$ and n$_{\rm H}$ given in Table 3 notes. For
the specific model adopted in this study the heating rate
scales according to $\Gamma_{x} \propto r^{-4.25}$ as 
illustrated in Figure 6.
As apparent from Figure 6 and Table 3,
the X-ray heating is mostly due to the cool component
as a result of its larger absorption cross-section and higher X-ray
luminosity (L$_{x,1}$). But for the disk  model adopted
here, X-ray ionization and heating are confined mainly to
upper layers several scale heights above the midplane (Figs. 4 and 5).
For the X-ray photon energies E $\ltsimeq$ 2.2 keV that
characterize the LkCa 15 {\em XMM-Newton} spectrum, 
gas surface densities $\Sigma_{0}$ $<$ 10$^{-2}$ g cm$^{-2}$
would be required in order for photons to penetrate
to deeper layers within one scale height of the midplane (Fig. 4).
Although such low surface densities were surmised in some early disk
models of LkCa 15 (A11b), more recent ALMA observations 
(van der Marel et al. 2015) suggest higher gas densities 
$\Sigma_{0}$ $\sim$ a few hundred g cm$^{-2}$

\clearpage
\begin{deluxetable}{cccccccc}
\tabletypesize{\footnotesize}
\tablewidth{0pt}
\tablecaption{X-ray Ionization and Heating Rates (LkCa 15)}
\tablehead{
           \colhead{r}               &
           \colhead{z/H$_{0}$}              &
           \colhead{E}           &
           \colhead{$\Sigma_{0}$} &
           \colhead{n$_{\rm H}$}         &
           \colhead{N$_{\perp}$}             &
           \colhead{$\zeta$}             &
           \colhead{$\Gamma_{x}$}      \\
           \colhead{(AU)}   &
           \colhead{}                 &
           \colhead{(keV)}               &
           \colhead{(g cm$^{-2}$)} &
           \colhead{(cm$^{-3}$)} &
           \colhead{(cm$^{-2}$)} &
           \colhead{(s$^{-1}$)}          &
           \colhead{(ergs s$^{-1}$ cm$^{-3}$)} 
 }
\startdata
  1 & 5.0 &  0.4 & 100 & 9.80e07   & 1.35e19 & 2.46e-09   & 6.56e-12  \\
  1 & 4.1 &  2.2 & 100 & 5.55e09   & 9.30e20 & 2.39e-11   & 3.61e-12  \\
\enddata
\tablecomments{
The ionization rate ($\zeta$) and heating rate ($\Gamma_{x}$)
for the cool (E = 0.4 keV) and hot (E = 2.2 keV) components are
based on the inner disk model discussed in the text (Sec. 5.1.1).
The rates are computed at $r$ = 1 AU for the specified value
$z$/H$_{0}$, which is the number of scale-heights above the midplane
corresponding to unit X-ray optical depth ($\tau_{x}$ = 1).
At $r$ = 1 AU the disk scale height is  
H$_{0}$ $\equiv$ H(r = 1 AU)  = 7.1 $\times$ 10$^{11}$ cm
for an assumed disk midplane temperature T($r$ = 1 AU) = 400 K.
A disk gas surface density 
$\Sigma_{0}$ $\equiv$ $\Sigma$(r = 1 AU) = 100 g cm$^{-2}$ is assumed,
but the actual value is uncertain and model-dependent.
The quantity n$_{\rm H}$ is the number density of H-nuclei
at the specified point ($r$,$z$) in the disk. 
The vertically-integrated H-nuclei column density from infinity down 
to the specified height above the midplane is N$_{\perp}$ (eq. [4]).
At $r$ = 1 AU, values at the midplane ($z$ = 0) are
n$_{\rm H}$($r$=1 AU,$z$=0) = 2.38 $\times$ 10$^{13}$ cm$^{-3}$ and
N$_{\perp}$($r$=1 AU,$z$=0) = 2.1 $\times$ 10$^{25}$ cm$^{-2}$. 
The primary ionization rates (eq. [2]) for the cool and hot plasma
components  are $\zeta_{x,1}$ = 2.53 $\times$ 10$^{-6}$ s$^{-1}$
and $\zeta_{x,2}$ = 4.31 $\times$ 10$^{-9}$ s$^{-1}$, respectively.
These values are computed using the average kT and L$_{x}$ values
for the two models in Table 2, namely
$\overline{\rm kT_{1}}$ = 0.4 keV,   $\overline{\rm kT_{2}}$ = 2.2 keV,
log $\overline{\rm L_{x,1}}$ = 30.26 ergs s$^{-1}$, and
log $\overline{\rm L_{x,2}}$ = 30.07 ergs s$^{-1}$.
The heating rate $\Gamma_{\rm x}$ = $\zeta$n$_{\rm H}$$Q$
(eq. [5]) is computed using Q = 17 eV.
\underline{Scaling Relations}:~H($r$) $\propto$ $r^{+1.5}$T($r$), or
H($r$) $\propto$ $r^{+1.25}$ for an assumed temperature
dependence T($r$) = 400($r$/1 AU)$^{-0.5}$. Scalings with radius
are $\Sigma \propto r^{-1}$, $n_{\rm H}$ $\propto$ $r^{-2.25}$,
$\zeta \propto r^{-2}$, $\Gamma_{x} \propto r^{-4.25}$.
At a given radius, 
$n_{\rm H}(r,z)$ = $n_{\rm H}(r,z=0$)exp[-0.5($z/H(r))^{2}$].
}
\end{deluxetable}
\clearpage

 \subsection{Comments on Circumstellar Disk Models}

Several different inner disk models have been proposed for LkCa 15
and the model adopted in our calculations above is only
representative. We have assumed 
$\Sigma(r)$ = $\Sigma_{0}$($r$/1 AU)$^q$ with $q$ = $-$1
and a gas surface density $\Sigma_{0}$ = 10$^{2}$ g cm$^{-2}$.
The adopted value of $\Sigma_{0}$ is similar to those determined in
recent studies but is about an order-of-magnitude less than values derived
in models of the minimum mass solar nebula (Weidenschilling 1977; Hayashi 1981).
By way of comparison, Manara et al. (2014) found 
$\Sigma$($r$ = 1 AU) = 83 g cm$^{-2}$ for LkCa 15.
Also, substitution of the LkCa 15 disk parameters determined in the ALMA
study of van der Marel et al. (2015) into the surface density profile
given in their eq. (1) yields $\Sigma$($r$ = 1 AU) = 286 g cm$^{-2}$.
This value includes their best-fit correction factor
$\delta_{gas}$ = 0.1 which scales down the gas density in the inner disk
(1 $<$ $r$ $<$ 45 AU) relative to the better-known gas distribution at 
larger radii.  But van der Marel et al. note that the 
value of $\delta_{gas}$ is only constrained to within an 
order-of-magnitude. Several factors contribute to its uncertainty
including inadequate spatial resolution to resolve gas in the inner disk, 
optical depth effects in the $^{12}$CO line used to map the disk gas
with ALMA, and possible asymmetries in the gas distribution. 
The much smaller surface density $\Sigma$($r$ = 1 AU) = 10$^{-3}$ g cm$^{-2}$
obtained previously by A11b reflects differences in observational data and
modeling strategy. The LkCa 15 disk model proposed in A11b was
based on 880 $\mu$m dust continuum observations obtained with
the Submillimeter Array (SMA) and assumed an empty cavity
devoid of dust and gas at radii 10 $\leq$ $r$ $\leq$ 50 AU.

The value of the gas surface density power-law exponent $q$  is
not well-constrained observationally in the inner disk but
$q$ = $-$1 is commonly used for LkCa 15 (e.g. Mulders et al. 2010;
Manara et al. 2014; van der Marel et al. 2015). But other surface
density profiles have been proposed such as the smooth viscous disk model
of Isella et al. (2012) which uses a positive value $q$ = $+$2.15
in the inner disk. In this model, the  surface density increases in
the inner disk and then starts to turn over at $r$ $\sim$ 60 AU
to match the observationally-constrained falloff in the outer
disk (Fig. 3 of Isella et al. 2012). The smooth viscous model
contrasts sharply with the discontinuous drop in 
$\Sigma(r)$ at $r$ $\sim$ 50 AU in the model of A11b.
The above differences are relevant
to  X-ray ionization and heating calculations
in the inner disk where the protoplanets are located
since the radial dependence of the density of H-nuclei  
is a function of the power-law index $q$ as n$_{\rm H}(r)$ $\propto$ $r^{q - 1.25}$
for an assumed  temperature dependence T($r$) $\propto$ $r^{-0.5}$.

We have adopted a disk temperature profile 
T($r$) = T$_{0}$($r$/1 AU)$^{\beta}$ K with the usual
exponent $\beta$ = $-$0.5 and normalized
to T$_{0}$ = 400 K for a stellar temperature
T$_{eff}$ = 4730 K. Other studies have used
T$_{0}$ = 100 K (A11a), 200 K (Manara et al. 2014),
$\sim$250 K (Pi\'{e}tu et al. 2007, scaled to
$r$= 1 AU), and 334 K (Bergin et al. 2004, scaled
to $r$= 1 AU and T$_{eff}$ = 4730 K). The temperature
inferred from the ALMA study of van der Marel et al. (2015)
is T$_{0}$ = 424 K, which is scaled from the dust sublimation radius
T($r$ = 0.08 AU) = 1500 K using $\beta$ = $-$0.5. 
This latter value is nearly identical to that adopted here.

The adopted disk temperature profile enters into X-ray ionization
and heating rate calculations through the scale height
H($r$) $\propto$ T($r$)$^{+0.5}$. This in turn
affects the mass-density $\rho$  and H-nuclei number density
n$_{\rm H}$ at the midplane
(Sec. 5.1.1). At a given radius, a higher disk
midplane temperature equates to a larger scale height H($r$)
and smaller n$_{\rm H}$. But the dependence is
weak and scales as n$_{\rm H}$($r$) $\propto$ T($r$)$^{-0.5}$,
so a factor of two uncertainty in T$_{0}$ translates into
only a factor of 1.4 uncertainty in n$_{\rm H}$ at $r$ = 1 AU.

\subsection{Comparison With Cosmic Ray Heating}

It is instructive to compare the X-ray heating rates in 
Table 3 with the rate expected from cosmic rays.
The cosmic ray heating rate for a gas consisting of 
H and H$_{2}$ is (Jonkheid et al. 2004) 

\begin{equation}
\Gamma_{cr} = \zeta_{cr}[2.5 \times 10^{-11}n_{\rm{H_{2}}} + 5.5 \times 10^{-12}n_{\rm{H}}] ~ (\rm{erg~ cm^{-3}~ s^{-1}})
\end{equation}
where the primary cosmic ray ionization rate 
in the interstellar medium (ISM)
is $\zeta_{cr}$ $\approx$ 5 $\times$ 10$^{-17}$ s$^{-1}$.
We assume that the  LkCa 15  disk is predominantly molecular hydrogen
and the second term in the above expression is negligible.
At $r$ = 1 AU the molecular hydrogen density at scale heights
corresponding to $\tau_{x}$ = 1 are half the n$_{\rm H}$
values given in Table 3, namely n$_{\rm{H_{2}}}$ = 4.9 $\times$ 10$^{7}$ cm$^{-3}$
($z$ = 5.0 H$_{0}$, 0.4 keV photons) and
2.8 $\times$ 10$^{9}$ cm$^{-3}$ ($z$ = 4.1 H$_{0}$, 2.2 keV photons).
We thus obtain $\Gamma_{cr}$($r$=1AU, $z$=5.0 H$_{0}$) = 
6.12 $\times$ 10$^{-20}$ ergs cm$^{-3}$ s$^{-1}$ and
$\Gamma_{cr}$($r$=1AU, $z$=4.1 H$_{0}$) =
3.48 $\times$ 10$^{-18}$ ergs cm$^{-3}$ s$^{-1}$. These values are
$\sim$10$^{6}$ - 10$^{8}$ times less than the X-ray heating
rates. Thus, in the upper disk layers X-ray heating dominates
and cosmic ray heating is negligible by comparison.
But at the midplane, X-ray heating  is
negligible for the surface density assumed here
($\Sigma_{0}$ = 10$^{2}$ gm cm$^{-2}$) whereas cosmic ray
heating is  $\Gamma_{cr}$($r$=1AU, $z$=0) = 
1.49 $\times$ 10$^{-14}$ ergs cm$^{-3}$ s$^{-1}$.

The winds and magnetic fields of T Tauri stars can impede
penetration of cosmic rays into the protoplanetary disk,
resulting in values of $\Gamma_{cr}$ that are less than the 
value appropriate for the ISM used above.  The effects of 
T Tauri star winds and magnetic fields on cosmic rays
were investigated by Cleeves et al. (2013) who concluded
that reduced primary  cosmic ray ionization rates of
$\Gamma_{cr}$ $\ltsimeq$ 10$^{-18}$ s$^{-1}$ are 
possible in protoplanetary disks, an order-of-magnitude
less than rates typically adopted for the ISM. The effect
of such lower cosmic ray ionization rates would be to
reduce cosmic ray heating near the disk midplane and
strengthen  the relative importance of X-ray heating
in the upper disk layers.

\section{X-ray Irradiation of the Protoplanets and Circumplanetary Disks}

The analysis above has focused on X-ray irradiation of the LkCa 15
circumstellar disk but an assessment of X-ray effects on the atmospheres
or circumplanetary disks of the protoplanets is also of interest.
At present, little is known about the atmospheric properties of the 
LkCa 15 protoplanets or their disks. But the detection of H$\alpha$ 
emission from the innermost protoplanet at $r$ $\approx$ 14.7 AU by 
Sallum et al. (2015) does provide evidence that it is accreting and 
a circumplanetary disk is likely present. However, an attempt to detect
7 mm continuum emission using the VLA within the inner disk gap where
the LkCa 15 protoplanets are located yielded negative results 
(Isella et al. 2014). Thus, crucial information on the temperature and
surface density distribution of the planetary atmospheres and
circumplanetary disks needed to calculate X-ray absorption and 
heating is lacking.

Nevertheless, we can show that if the LkCa 15 circumstellar disk
surface density at a distance $r$ = 1 AU from the star is of order
$\Sigma_{0}$ $\sim$ 10$^{2}$ g cm$^{-2}$, as assumed in our calculations above,
then little if any stellar X-ray emission reaches  protoplanets 
that are near the midplane ($z$ $\approx$ 0) at distances of
$r$ $\sim$ 15 - 20 AU. At a distance $r$ from the star
the intrinsic  stellar X-ray flux F$_{x}$ is
reduced to F$_{x}$e$^{-\tau_{x}}$/$r^2$ where
where $\tau_{x}$ $\rightarrow$ $\tau_{x}$($r,z,$E) is the energy-dependent 
X-ray optical depth through the circumstellar disk evaluated at the target 
point ($r,z$). Assuming as before $\Sigma(r)$ = $\Sigma_{0}$($r$/1 AU)$^{-1}$
and $\Sigma_{0}$ = 10$^{2}$ g cm$^{-2}$ then we obtain by Equation (4)  
N$_{\perp,{\rm disk}}$(r=15 AU,z=0) = 1.4 $\times$ 10$^{24}$ cm$^{-2}$.
Using Equation (3) the optical depths for the cool and hot plasma components
are computed to be  $\tau_{x}$($r$=15 AU,$z$=0,E=0.4 keV) = 1.6 $\times$ 10$^{6}$
and $\tau_{x}$($r$=15 AU,$z$=0,E=2.2 keV) = 2.3 $\times$ 10$^{4}$.
T Tauri stars may undergo intermittent bright X-ray flares during which plasma 
temperatures can briefly reach  peak temperatures T $\sim$ 100 - 250 MK (E $\sim$ 9 - 22 keV)
as has been well-documented in the Taurus Molecular Cloud (Franciosini et al. 2007)
and Orion Nebula Cluster (Favata et al. 2005). No such high-temperature flares 
have been detected in LkCa 15 so far but optical depth calculations for the above
range of peak values give $\tau_{x}$($r$=15 AU,$z$=0,E=9 keV) = 685 and
$\tau_{x}$($r$=15 AU,$z$=0,E=22 keV) = 74. At these large optical optical depths
it is obvious that any protoplanets located in or near the circumstellar disk midplane
at $r$ $\gtsimeq$ 15 AU will be shielded quite effectively from direct stellar X-ray emission.

The main uncertainty in the above $\tau_{x}$ calculations originates
in the assumed surface density profile $\Sigma(r)$ = $\Sigma_{0}$($r$/1 AU)$^{q}$.
As already mentioned (Sec. 5.2), the normalization $\Sigma_{0}$ and 
power-law exponent $q$ are not well-constrained by observations for LkCa 15,
and the dependence of $\Sigma(r)$ with radius could be more complex
than a simple power-law. But in order for 
stellar X-rays to penetrate through the circumstellar disk midplane 
to reach the protoplanets, much lower surface densities than assumed
above would be required. Since $\tau_{x}$ $\propto$ $\Sigma_{0}$
for fixed ($r,z,$E), values $\Sigma_{0}$ $\sim$ 10$^{-3}$ g cm$^{-2}$
(as in the model of A11b; see also SG13) would be needed to achieve 
$\tau_{x}$($r$=15 AU,$z$=0,E) $\sim$ 1.
If the gas density is not currently that low it will eventually
become so as the star-disk system evolves and the inner-disk gas dissipates.

\section{Summary}

A pointed observation of LkCa 15 with {\em XMM-Newton}
confirms that it is a luminous X-ray source, as previously
inferred from a shorter {\em Chandra} observation.
Although some low-level count-rate variability is likely 
present in the  {\em XMM-Newton} X-ray light curves,
no large-amplitude variability or flares were detected.
The X-ray spectrum is characterized by a two-temperature
thermal plasma with temperatures kT$_{cool}$ $\approx$ 0.4 keV 
and a hotter  component at kT$_{hot}$ $\approx$ 2.2 keV. 
The temperature of the hot component is about half that
determined from the {\em Chandra} spectrum, suggesting
that the hot component is variable as is often the
case in T Tauri stars. 
Most of the X-ray ionization and heating of the  
circumstellar disk is due to the cool component.
X-ray heating is restricted mainly to the upper disk
layers assuming a disk gas surface density at $r$ = 1 AU
of $\Sigma_{0}$ $\sim$ 10$^{2}$ g cm$^{-2}$, as suggested by recent
ALMA observations. Cosmic-ray heating is negligible compared
to X-ray heating in the upper disk layers but few X-ray
photons reach the disk midplane, where cosmic ray heating 
dominates. Unless the disk gas surface density is  much
less than the assumed value $\Sigma_{0}$ $\sim$ 10$^{2}$ g cm$^{-2}$,
little or no stellar X-ray emission is able to 
penetrate the circumstellar disk midplane to distances of
$r$ $\approx$ 15 - 20 AU where the protoplanets are located.

\acknowledgments

This work was supported by NASA Astrophysics Data Analysis
Program (ADAP)  award NNX16AL71G.
This work  was based on observations obtained with
{\em XMM-Newton}, an ESA science mission with instruments and
contributions directly funded by ESA member states and
the USA (NASA). This research has made use of the HEASOFT
data analysis software develeoped and maintained by HEASARC
at NASA GSFC.

%
%


\clearpage





\begin{figure}
\figurenum{1}
\epsscale{1.0}
\includegraphics*[width=10.0cm,angle=0]{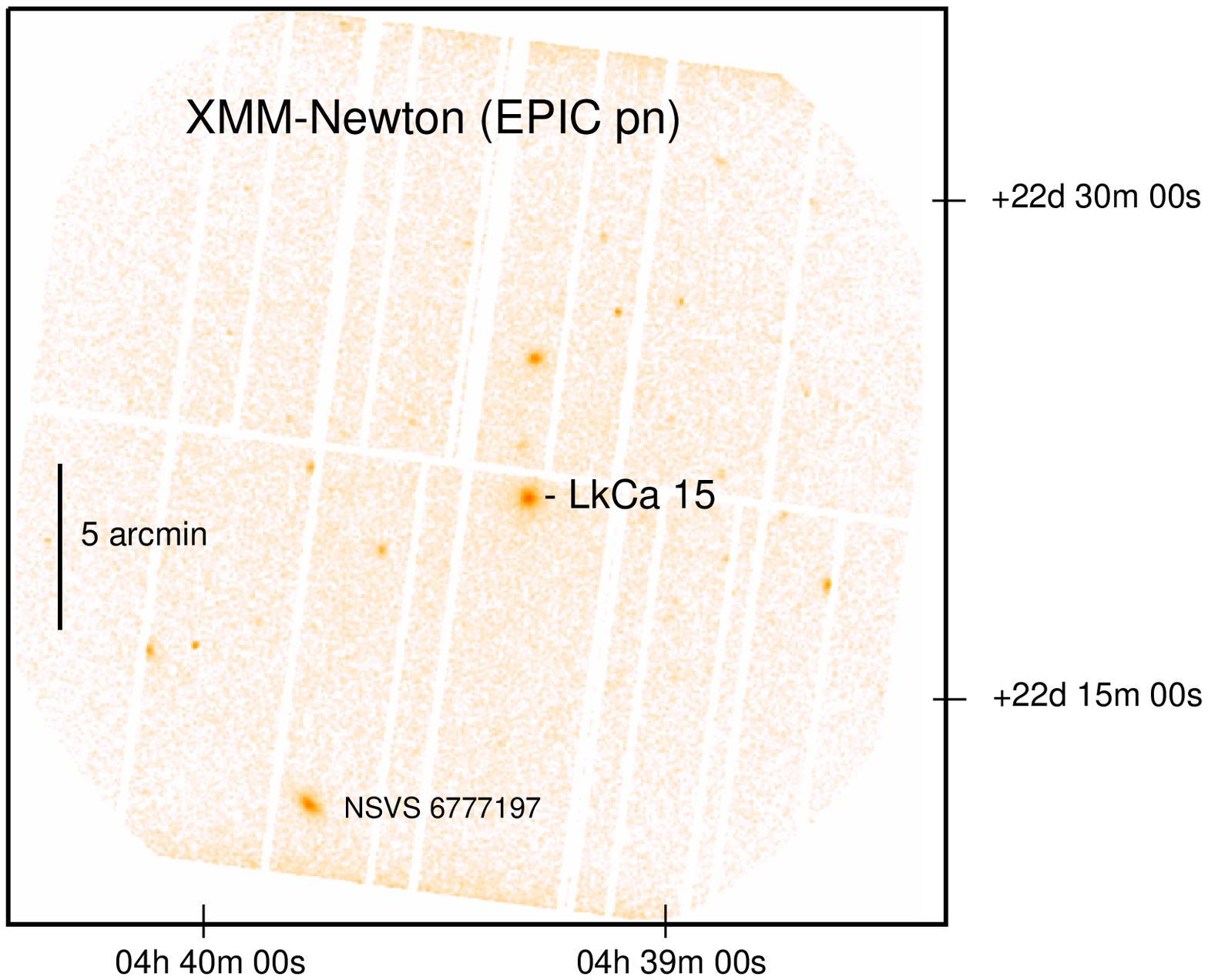}

\caption{Broad-band (0.2 - 8 keV) lightly-smoothed EPIC pn image of LkCa 15
and surrounding region. The image has been time-filtered to 
remove intervals affected by background flares. The Algol-type
binary NSVS 6777196 is marked near the south edge of the pn 
field-of-view. The bright source 4.2$'$ north of LkCa 15
is classified as non-stellar in the {\em HST} v. 2.3
Guide Star Catalog (GSC J043916.97$+$222515.59).
The image is displayed on a log intensity scale with N up
and  E to left. The coordinates are equinox J2000.
}
\end{figure}

\clearpage

\begin{figure}
\figurenum{2}
\epsscale{1.0}
\includegraphics*[width=8.0cm,angle=-90]{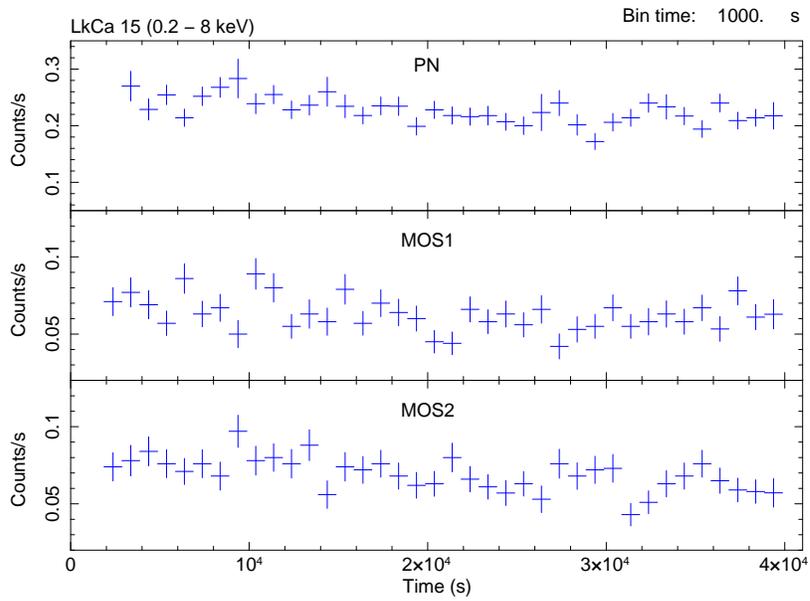} 
\caption{
Background-subtracted EPIC light curves of LkCa 15 in the 0.2-8 keV range
extracted from a circular region of radius 20$''$ centered on the source.
Binned at 1000 s intervals. Intervals of high background have
been removed. Times are relative to 19:00 UTC on 2014 Feb. 20.
}
\end{figure}

\clearpage

\begin{figure}
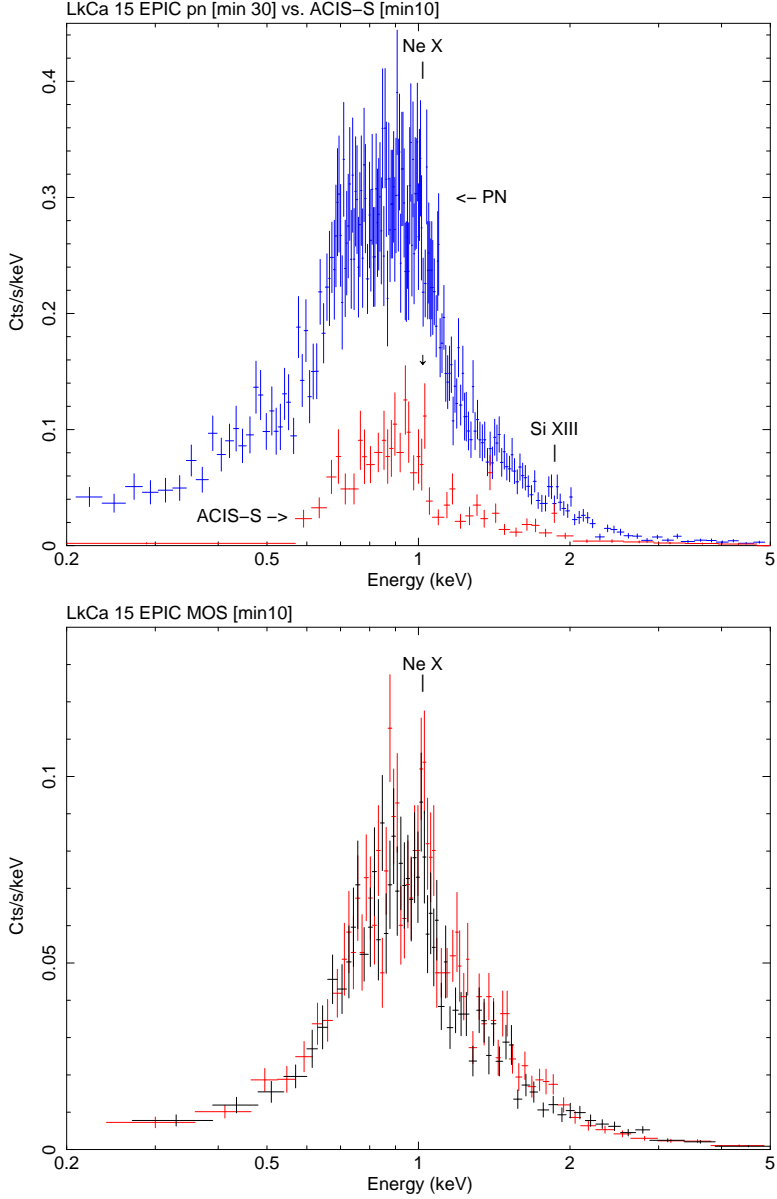

\figurenum{3}
\epsscale{1.0}
\includegraphics*[width=8.0cm,angle=-90]{f3t.eps} \\
\includegraphics*[width=8.0cm,angle=-90]{f3b.eps}
\caption{
Background-subtracted X-ray spectra of LkCa 15 obtained with {\em XMM-Newton} EPIC 
(ObsId 0722340101) and  {\em Chandra} ACIS-S (ObsId 10999). 
Intervals of high-background emission have been
excluded from the EPIC spectra. ACIS-S background is negligible.
~Top:~EPIC pn (blue, 7568 net counts, binned to a minimum of 30 counts per bin) overlaid with 
ACIS-S (red, 590 net counts, binned to a minimum of 10 counts per bin). 
Possible Ne X (1.02 keV) and Si XIII (1.86 keV) lines are marked.~
Bottom:~Overlay of  EPIC  MOS1 (black; 2463 net counts) and MOS2 (red; 2743 net counts)  
spectra, binned to a minimum of 10 counts per bin. A spectral feature visible in 
both MOS1 and MOS2 identified as Ne X (1.02 keV) is marked. 
}
\end{figure}

\clearpage

\begin{figure}
\figurenum{4}
\epsscale{1.0}
\includegraphics*[width=10.5cm,angle=-90]{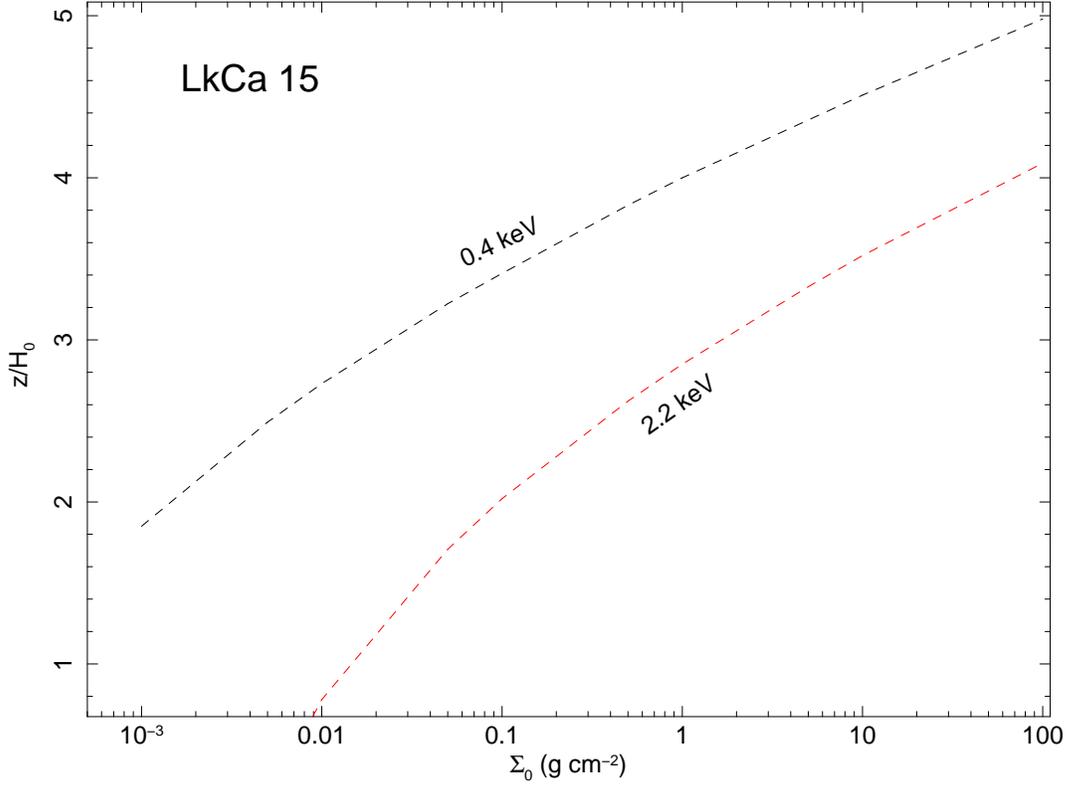}
\caption{Distance above the disk midplane corresponding to unit
X-ray optical depth ($\tau_{x}$ = 1)  for the cool and hot plasma
components as a function of the  disk gas surface density $\Sigma_{0}$ at $r$ = 1 AU. 
The height above the midplane is expressed in units of the 
scale height H$_{0}$ $\equiv$ H($r$ = 1 AU) = 7.1 $\times$ 10$^{11}$ cm
for an assumed disk midplane temperature T($r$ = 1 AU) = 400 K.
Very low surface densities $\Sigma_{0}$ $<$ 0.01 g cm$^{-2}$
are required for the X-rays to penetrate to depths below
one scale height.
}
\end{figure}

\clearpage

\begin{figure}
\figurenum{5}
\epsscale{1.0}
\includegraphics*[width=10.5cm,angle=-90]{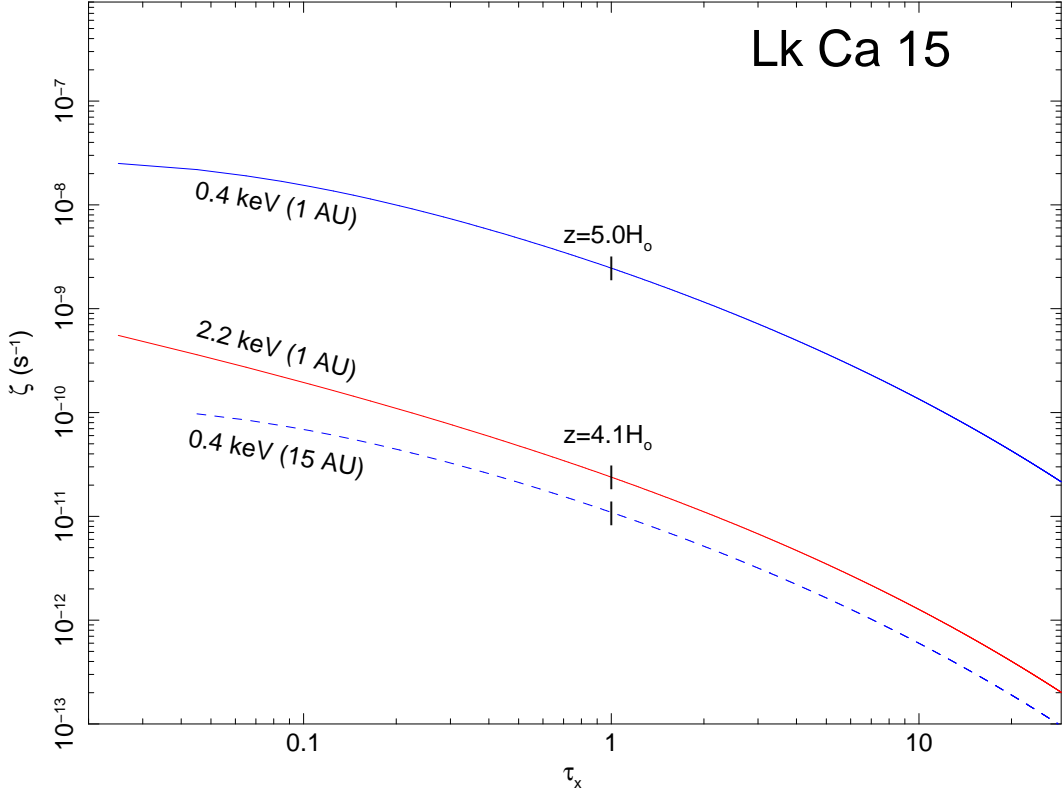}
\caption{X-ray ionization rate versus  X-ray optical depth for the LkCa 15 disk
as computed for the model adopted in this study (Sec. 5; Table 3). The two solid curves 
show the ionization rates for the cool and hot components evaluated at a radial distance
r = 1 AU from the star. The dashed curve shows the rate for the cool component at r = 15 AU. 
The short vertical lines mark the scale height corresponding to $\tau_{x}$ = 1
assuming a disk gas surface density $\Sigma_{0}$ $\equiv$ $\Sigma$($r$=1 AU) =
10$^{2}$ g cm$^{-2}$. At $r$ = 1 AU the scale height is 
H$_{0}$ = 7.1 $\times$ 10$^{11}$ cm (= 0.048 AU) for an assumed disk
temperature T($r$=1 AU) = 400 K.
The ionization rate for a specific height $z$ scales as 
$\zeta(r) \propto r^{-2}$. 
}
\end{figure}

\clearpage

\begin{figure}
\figurenum{6}
\epsscale{1.0}
\includegraphics*[width=10.5cm,angle=-90]{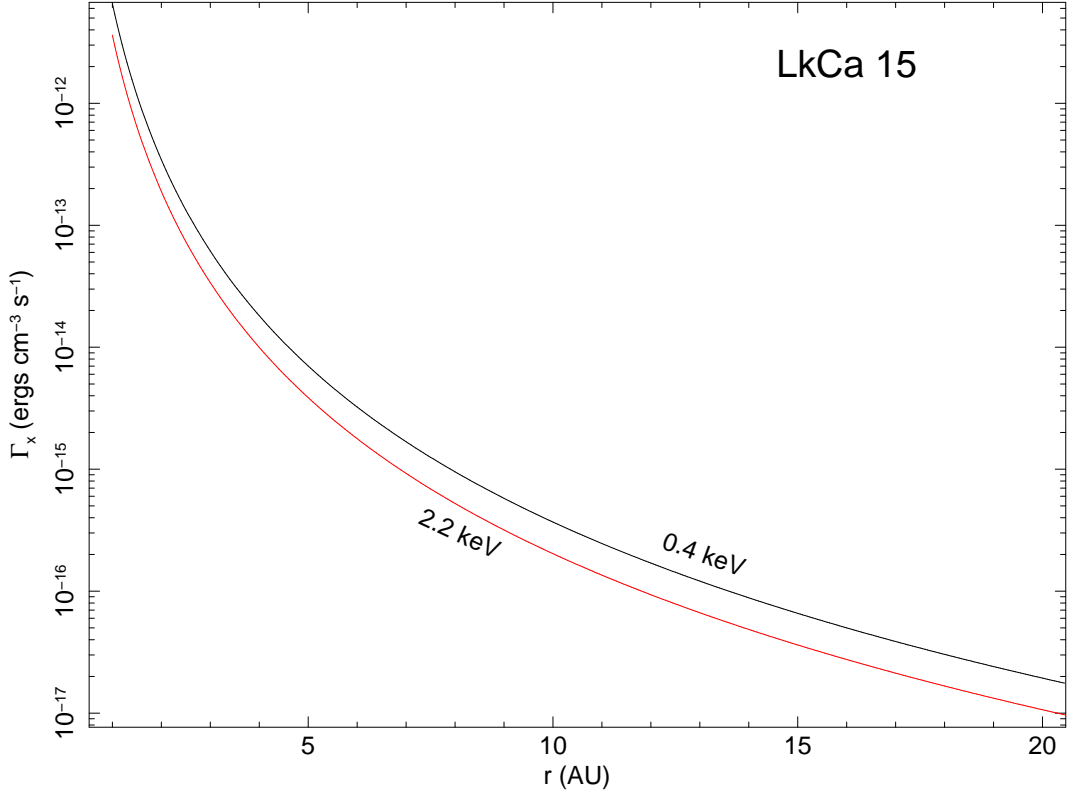}
\caption{Disk heating rate per unit volume as a function of radial
distance from the star for the cool and hot plasma components.
The heating rate is evaluated at the distance above the midplane
corresponding to $\tau_{x}$ = 1 (see Fig. 4), which is
z/H$_{0}$ = 5.0 (0.4 keV) and z/H$_{0}$ = 4.1 (2.2 keV).
For the disk model adopted in this study $\Gamma_{x}$ $\propto$ $r^{-4.25}$.
} 
\end{figure}

\end{document}